\documentclass[twocolumn,preprintnumbers,amsmath,amssymb,superscriptaddress]{revtex4-1}
\usepackage{amsmath}
\usepackage{graphicx}
\usepackage{dcolumn}
\usepackage{bm}
\usepackage{natbib}

\usepackage{color}
\definecolor{red}{rgb}{1,0,0}

\definecolor{blue}{rgb}{0,0,1}

\definecolor{green}{rgb}{0,1,0}

\setcitestyle{square}

\begin{document}
\preprint{APS}
\author {Gabriel Gomes}
\affiliation{S\~ao Paulo University (USP), IAG, Departamento de
  Astronomia, S\~ao Paulo, SP, Brazil}
\author {H. Eugene Stanley}
\affiliation{Center for Polymer Studies and Department of Physics,
  Boston University, Boston, Massachusetts 02215, USA}
\author {Mariano de Souza}
\affiliation{S\~ao Paulo State University (UNESP), IGCE, Departamento de
  F\'isica, Rio Claro, SP, Brazil}

\title{Enhanced Gr\"uneisen Parameter in Supercooled Water}

\begin{abstract}

\noindent
We use the recently-proposed \emph{compressible cell\/} Ising-like model
[Phys. Rev. Lett. \textbf{120}, 120603 (2018)] to estimate the ratio
between thermal expansivity and specific heat (the Gr\"uneisen parameter
$\Gamma$) in supercooled water. Near the critical pressure and
temperature, $\Gamma$ increases. The $\Gamma$ value diverges near the
pressure-induced finite-$T$ critical end-point
[Phys. Rev. Lett. \textbf{104}, 245701 (2010)] and quantum critical
points [Phys. Rev. Lett. \textbf{91}, 066404 (2003)], which indicates
that two energy scales are governing the system. This enhanced behavior
of $\Gamma$ is caused by the coexistence of high- and low-density
liquids [Science \textbf{358}, 1543 (2017)]. Our findings support the
proposed liquid-liquid critical point in supercooled water in the
No-Man's Land regime, and indicates possible applications of this model
to other systems.

\end{abstract}

\maketitle

\begin{figure}[htb]
\centering
\includegraphics[width=0.9\columnwidth]{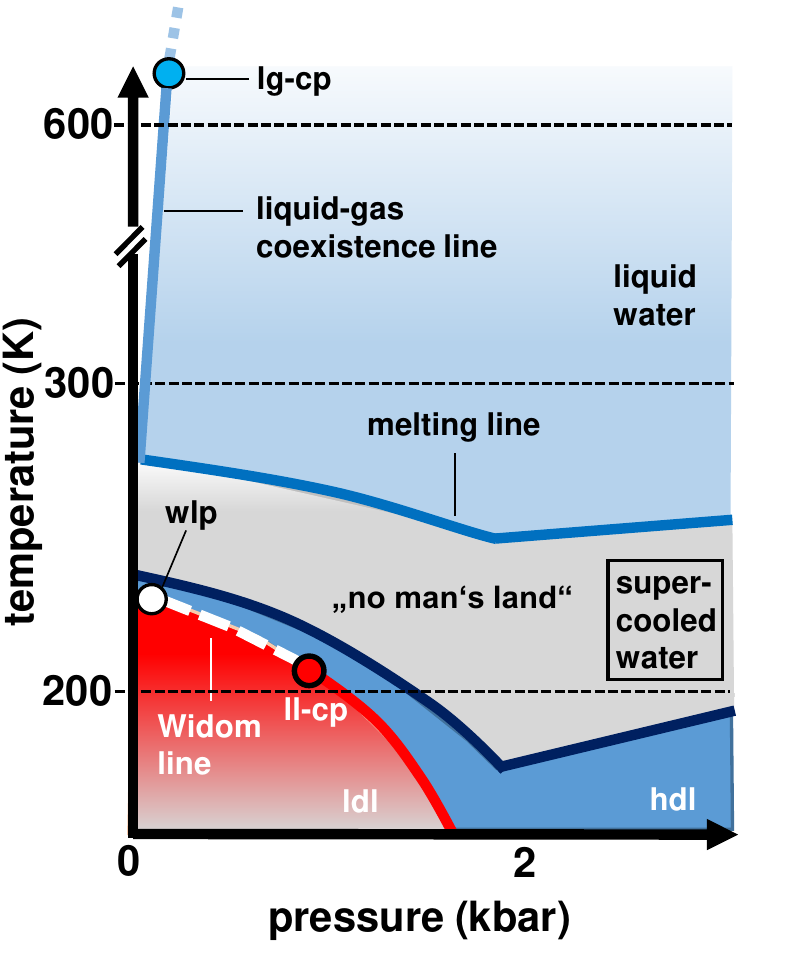}
\caption{Temperature \emph{versus} pressure phase diagram of water,
  lg-cp refers to liquid-gas critical point, wlp is the Widom line point
  and ll-cp indicates the liquid-liquid critical point, which is the
  focus of the present work. Picture after
  \cite{Gallo1543,Gallo}.}\label{Fig-1}
\end{figure}

\noindent
Because it is biologically fundamental in the maintenance of all life,
liquid water is one of the most important substances on the
planet. Water exhibits a number of anomalous physical properties (see
Fig.~\ref{Fig-1} and Ref.~\cite{PhysicsToday,Gallo}), and over the last
25 years much water research has focused on its so-called supercooled
phase. The initial work on supercooled water in 1992 used molecular
dynamic simulations \cite{Poole} and subsequent research has explored
the No-Man's Land region in the phase diagram (see Fig.~\ref{Fig-1} and
Ref.~\cite{Gallo}). This topic has generated much debate
(\cite{Kumar9575,Franzese,Gallo,Gallo1543,Kim1589} and references
therein).

One scenario describing supercooled water assumes the existence of two
liquid phases at low-$T$, one that is high-density, the other
low-density \cite{Gallo1543}. Recently fs x-ray scattering was used on
water droplets to determine the maximum isothermal compressibility, the
correlation length, and the structures of water and
heavy water. Experimental evidence of a second-order critical end-point
in the Widom line was found \cite{Kim1589}, but no clear-cut divergence
in the quantities was observed. Here we use the Gr\"uneisen parameter
($\Gamma$) \cite{Barto,2015,EPJ} on supercooled water and find evidence
supporting a liquid-liquid critical point. We use a recently-proposed
\emph{compressible\/} Ising-like model \cite{claudio,Cerde,Cerde1} to
obtain $\Gamma$. The model assumes two volumes $v_0$ and $v_0 + \delta
v$, with $\delta v > 0$. The two characteristic volumes are $0 <
\dot{v}_+ < v_+$ and $0 < \dot{v}_- < v_-$, and their ratio is
$\lambda$, where $\lambda = \dot{v}_+ / \dot{v}_-$.

The system has $N$ sites and coordination number $c$. Because we assign
a cell to each site, particles can move through the free volume.  The
interaction between sites is $\delta \varepsilon$, the system energy $E
\{n_i\}$ is
\begin{equation}
E \{n_i\} = \frac{cN\varepsilon _0}{2} - \delta \varepsilon \sum _{<ij>}
n_i n_j,
\label{Hamiltonian}
\end{equation}
where $\varepsilon _0$ is an arbitrary energy value, the volume is \cite{claudio}
\begin{equation}
 V \{n_i\} =  v_0 \sum _{i,j} (n_i + n_j) + \sum _j n_j \delta v = v_0 N
 + \delta v \sum _{\
i=1} ^N n_i,
\label{Volume}
\end{equation}
where $(n_i + n_j) = N$, $\{n_i\}$ when $K$ particles occupy volume
$v_-$ and $(N-K)$ particles occupy $v_+$.  Thus each particle is located
in a site, and the volume has two possible values. We associate these
two volumes with the low- and high-density phases and thus with two
distinct energy scales. These are key in understanding why $\Gamma$ is
enhanced near the liquid-liquid critical point. We obtain all the
observables related to the system from Eqs.~(\ref{Hamiltonian}) and
(\ref{Volume}) \cite{huang}. We carry out an isothermal-isobaric
analysis and sum $e^{-E/k_B T}$ and $e^{-pV/k_B T}$ to the partition
function, where $k_B$ is the Boltzmann constant and $p$ and $T$ are the
pressure and temperature of all possible microstates of the system,
respectively.

The resulting partition function $Z = Z(N,p,T)$ has the same
mathematical structure as the Ising canonical partition
function. Because we have not yet solved the three-dimensional Ising
model, we use an approximate \textit{mean-field solution\/}
\cite{claudio} to obtain the observables. The mean-field theory can be
applied to a wide range of systems, including the Ising model and the
van der Waals theory for liquid-gas systems \cite{huang}. Using it we
replace the functional integral $Z = N \int (Dm) e^{-E[m,H]}$ with the
maximum value of the integrand, the so-called \textit{saddle-point
  approximation}. The parameter $m$ is the order-parameter density, and
$Dm$ is the volume element. Because this approximation assumes that the
only important configuration near the critical point is the one of
uniform density, we expect that, because the density fluctuations in the
order parameter are strong in this regime, this study of critical
phenomena will exhibit artifacts. However Ref.~\cite{claudio} indicates
that consistent results can be obtained in this framework. The equation
of state for the system is \cite{claudio}
\begin{equation}
p(T,v) = \frac{Tk_B}{\delta v} \ln\left( \lambda \frac{v_0 + \delta v -
  v}{v - v_0} \right) + c \frac{\delta \varepsilon}{\delta v}
\frac{v-v_0}{\delta v},
\label{PVT}
\end{equation}
from which we deduce
\begin{equation}
T(p,v) = \frac{\delta v}{k_B f(v)} \left[ p - c \frac{\delta
    \varepsilon}{\delta v ^2} (v-v_0) \right].
\label{Temperature}
\end{equation}
We use Eq.~(\ref{PVT}) to determine the critical point coordinates $p_c
= (v_c,T_c)$ \cite{huang}
\begin{equation}
\left(\frac{\partial p}{\partial v} \right) _T = 0;\hspace{1cm}
\left(\frac{\partial ^2 p}{\partial v ^2} \right) _T = 0.
\label{condition_1}
\end{equation}
Thus
\begin{equation}
\left( \frac{\partial p}{\partial v} \right) _T = \frac{c \delta
  \varepsilon}{\delta v ^2} - \frac{T k_B}{(v_0 + \delta v - v)
  (v-v_0)},
\end{equation}
and
\begin{equation}
\left( \frac{\partial ^2 p}{\partial v ^2} \right) _T = Tk_B \left[
  \frac{2(v_0-v)+\delta v}{(v_0 + \delta v - v)^2 (v-v_0)^2} \right].
\end{equation}
We apply these conditions and the critical point parameters are
$$ v_c = v_0 + \frac{1}{2} \delta v; \hspace{0.8cm} T_c = \frac{c \delta
  \varepsilon}{4 k_B}, \hspace{0.8cm} p_c = \frac{c}{4} \frac{\delta
  \varepsilon}{\delta v} (2+\ln \lambda).$$

\begin{figure}
\centering
\includegraphics[width=\columnwidth]{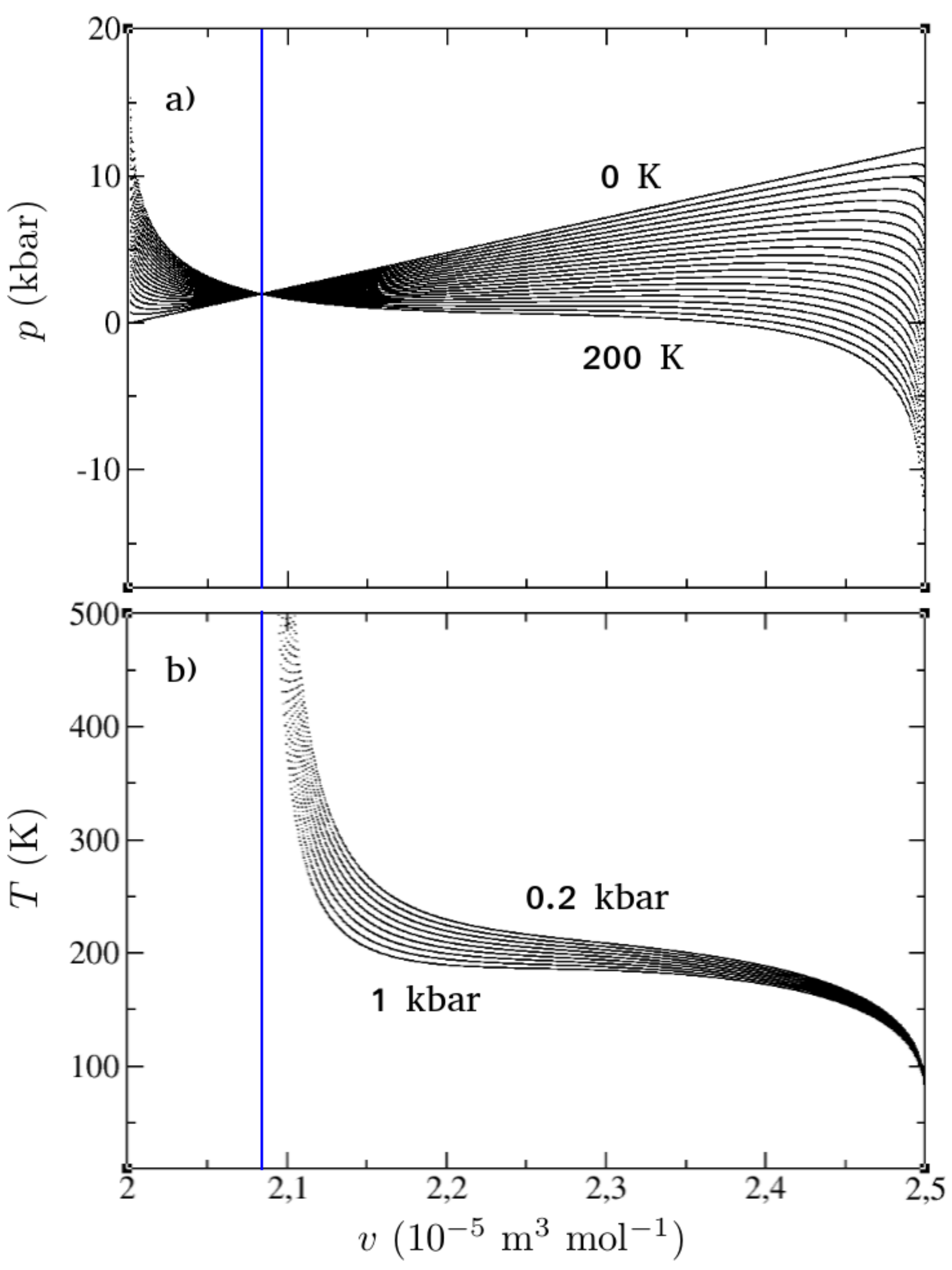}
\caption{a) Pressure ($p$) \emph{versus} volume ($V$) phase diagram
  obtained from Eq.\,(\ref{PVT}) for different values of temperature. The
  temperature was uniformly varied from 0 to 200\,K, with steps of
  10\,K. The straight line is related to $T = 0$\,K.  Similar results
  were reported in Ref.\,\citep{claudio}. b) Temperature ($T$)
  \emph{versus} volume ($V$) for different values of pressure, which
  were also varied uniformly as in panel a). The parameters used were
  the same as in \citep{claudio}, namely $c = 6$, $\delta \varepsilon =
  1000$ J mol$^{-1}$, $v_0 =(2 \times 10^{-5})$ m$^3$ mol$^{-1}$,
  $\delta v = (0.5 \times 10^{-5})$ m$^3$ mol$^{-1}$ and $\lambda =
  0.2$.}
\label{Fig-2}
\end{figure}

Employing the basic thermodynamic relations \citep{huang} and using
$f(v) = \ln\left( \lambda \frac{v_0 + \delta v - v}{v - v_0} \right)$
\cite{claudio} we obtain the isobaric thermal expansivity $\alpha _p$
and the heat capacity $c_p$,
\begin{equation}
\alpha _p = \frac{1}{v} \left[ \frac{\delta v ^2}{k_B f(v)^2} g(v) -
  \frac{c \delta \varepsilon}{k_B \delta v f(v)} \right] ^{-1},
\label{alpha}
\end{equation}
$$ g(v) = \frac{1}{(v_0 + \delta v - v) (v - v_0)},$$
and
\begin{equation}
c_p = T \frac{k_B}{\delta v} f(v) \left[ \frac{\delta v ^2}{k_B f(v)^2}
  g(v) - \frac{c \delta \varepsilon}{k_B \delta v f(v)} \right] ^{-1}.
\label{Cp}
\end{equation}
We use Eqs.~(\ref{alpha}) and (\ref{Cp}) to determine the expression of
$\Gamma$ for the system. By definition $\Gamma = \frac{\alpha _p}{c_p}$
\cite{EPJ}, and thus using Eqs.~(\ref{alpha}) and (\ref{Cp}) we have
\begin{equation}
\Gamma = \frac{\delta v}{T k_B v} \ln\left( \lambda \frac{v_0 + \delta v
  - v}{v - v_0} \right).\label{Grueneisen}
\end{equation}

\begin{figure}
\centering
\includegraphics[width=\columnwidth]{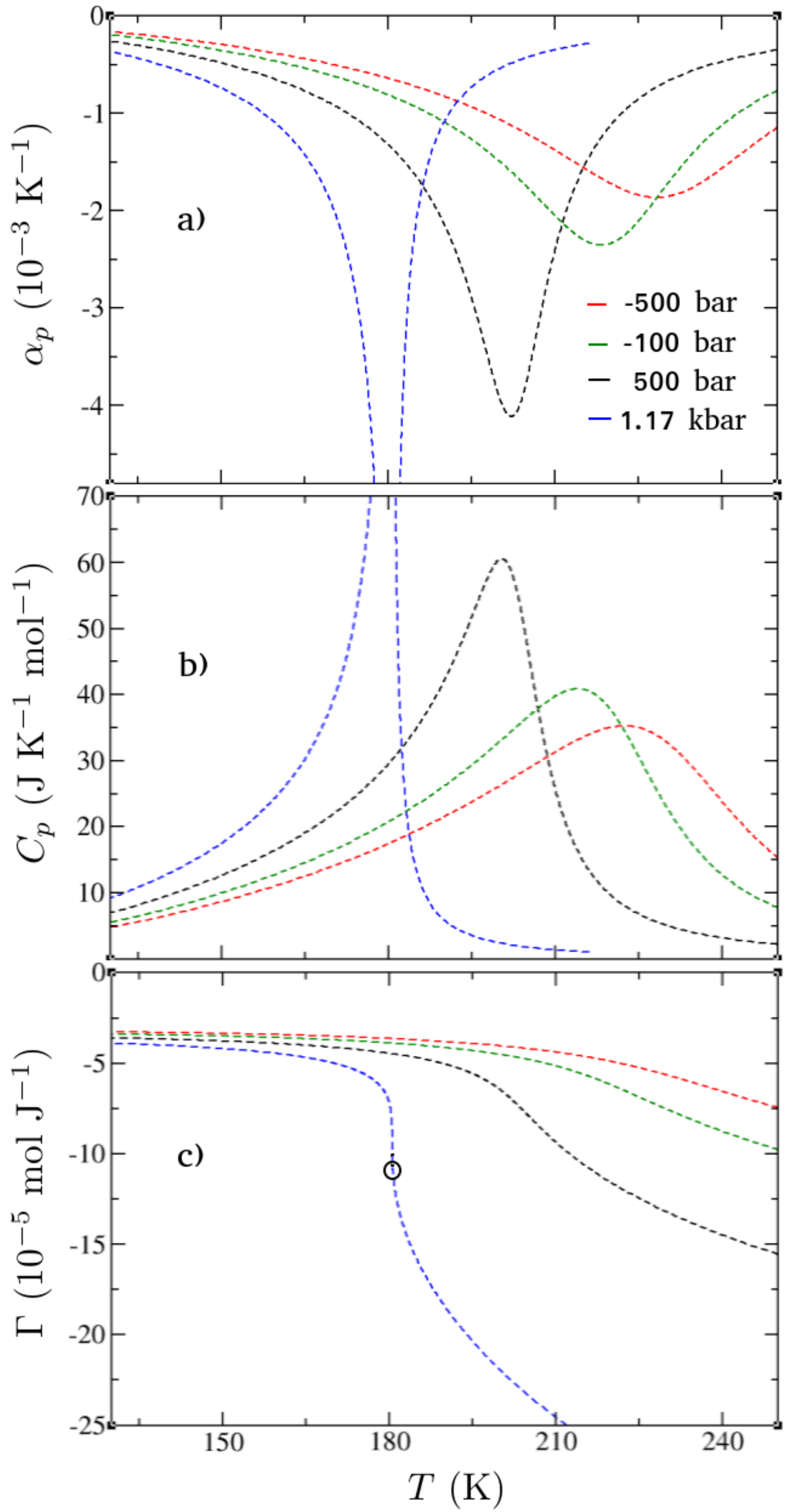}
\caption{a) Isobaric thermal expansivity $\alpha_p$, b) isobaric heat
  capacity $c_p$ and c) Gr\"uneisen parameter $\Gamma$ for different
  values of pressure. The employed parameters were the same as presented
  in the caption of Fig.\,\ref{Fig-2}. The critical point is indicated
  by the open circle in c). Further details are discussed in the main
  text.}
\label{Fig-3}
\end{figure}

We have obtained all the observables and now analyze their behavior near
the critical point. Note that because the above equations take the form
$c_p [T(p,v),]$, $\alpha _p (v)$, and $T(p,v)$ they constitute a
parametric system. This prevents our obtaining an analytical expression
for $v$, and Eq.~(\ref{PVT}) clearly indicates a transcendental equation
in $v$. We thus analyze the behavior of $c_p$ and $\alpha _p$ by varying
$v$, which causes variations in $T$. We fix the critical point
parameters by employing the corresponding expressions. We adjust the
parameters \cite{claudio} so that $T_c \simeq 180$\,K, which is in the
No Man's Land region \cite{Gallo}.

Figures \ref{Fig-2}(a) and \ref{Fig-2}(b) show the $p$--$V$ phase
diagram for a range of temperatures and the $T$--$V$ diagram. Note that
when $T = 0$K the resulting mapping $p(0, v)$ is a straight line. This
is obtained using Eq.~(\ref{PVT}). When the temperature high, the
pressure for $v \approx v_0$ is higher than when the temperature is
low. For $v \approx v_0 + \delta v$, however, higher temperatures
decrease the pressure for fixed values of $v$. Figure~\ref{Fig-2}(b)
shows that in a particular range of values of volume, given the pressure
values, physical temperature values are
inaccessible. Figure~\ref{Fig-2}(a) shows that the point where the
pressure is the same for every temperature value (blue vertical line) is
the limiting value for the volume $(V)$ for which physical values of the
temperature are obtained. Figure~\ref{Fig-2}(b) shows the results using
Eq.~(\ref{PVT}). Note that we cannot analytically obtain expression
$v(T,p)$ because Eq.~(\ref{PVT}) is transcendental in $V$. Thus we have
a mapping of these physical quantities [see Eq.~(\ref{Temperature})],
and we can find the corresponding $v$ and $T$ values for each pressure
value $(p)$. The same holds for any other desired order of these three
parameters. Figures~\ref{Fig-3}(a), \ref{Fig-3}(b), and \ref{Fig-3}(c)
show the behaviors of $\alpha _p$, $c_p$, and $\Gamma$, respectively,
for different values of pressure. Note that as the pressure is
increased, both the minimum of $\alpha _p$ and the maximum of $c_p$
shift to lower values of temperature, and $\Gamma$ becomes steeper. These
features are caused by their proximity to the critical point, which is
$p_c \approx 1.17$\,kbar \cite{claudio}.

Figure \ref{Fig-3}(c) shows the effect of pressure on $\Gamma$ and a
distinct behavior upon approaching the critical point. Note that the
$c_p$ and $\alpha _p$ features for pressure values near $p_c$ are
distinct from those observed for $\Gamma$. For $\Gamma$, as $p_c$ is
approached the variation with temperature reaches a maximum at $T =
T_c$, indicating the presence of a critical point
\cite{Barto,EPJ,garst,Cerio}. This is our key finding.

The Maxwell-relation $\left(\frac{\partial V}{\partial T} \right)_p = -
\left(\frac{\partial S}{\partial p} \right)_T$ and the negative thermal
expansivity shown in Fig.~\ref{Fig-3} indicate that the entropy of the
system is enhanced when approaching the liquid-liquid critical point,
i.e., by applying pressure the high- and low-density phases mix and the
entropy increases. We also find this in the finite-$T$ critical
end-point reported for molecular conductors \cite{Barto,Review-kappa}
and the quantum critical points in heavy-fermion compounds
\cite{Zhu,Kuchler}. The high- and low-density phases produce two
different energy scales. Because the degree of H-bonding depends on
temperature and pressure, a scaling cannot be applied successfully
\cite{Roland, Cook}.  Reference~\cite{Gallo1543} indicates that water
molecule interactions create an open H-bond structure that has a lower
density than other configurations. We can capture the energy-scales
associated with H-bond configurations that correspond to the low- and
high-density phases using a compressible Ising-like model and two
accessible system volumes. Using the Landau theory \cite{Kadanoff} we
find that decreasing the order-parameter fluctuations creates
divergences in the correlation length \cite{Kim1589} and relaxation time
\cite{Relaxation-time}. Reference~\cite{Casalini} reports a connection
between the entropy-dependent relaxation time and $\Gamma$. We here
suggest that this also is true for supercooled water.

\begin{figure}
\centering
\includegraphics[width=\columnwidth]{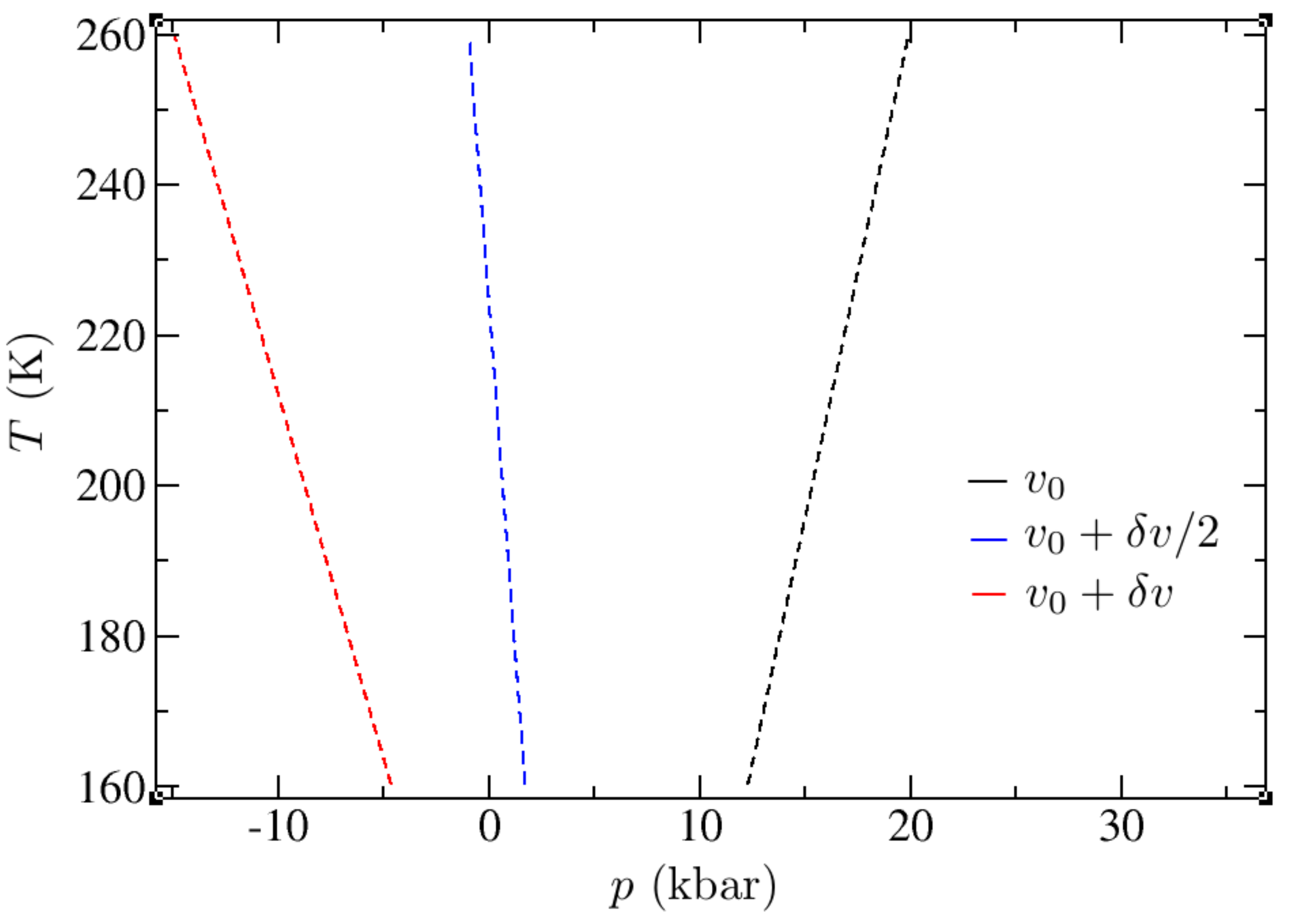}
\caption{Temperature ($T$) \emph{versus} pressure ($p$) phase diagram for
  three different values of volume ($v$), as indicated in the label. A
  linear relation is observed between $T$ and $p$ for all values of
  $v$. Moreover, for values near the superior limit of the volume, the
  pressure reaches negative values and the angular coefficient of the
  mathematical relation between $T$ and $P$ changes sign. The values of
  $v_0$ and $\delta v$ employed here are the same as in
  Fig.\,\ref{Fig-2}.}
\label{Fig-4}
\end{figure}

We have studied the thermodynamic quantities of a given set of fixed
initial parameters. We now extend the model and use it to theoretically
predict critical behaviour. Here parameter $\lambda$ is the ratio of
characteristic volumes that each particle can occupy, and by changing
the $\lambda$ ratio we can use the model to simulate other systems. We
thus next analyze $\Gamma$ by varying $\lambda$.

We analyze Eq.~\ref{Grueneisen} and find that varying $\lambda$ changes
the critical value of $p_c$, but that $T_c$ remains the same. This is
caused by the non-dependence of $T_c$ on $\lambda$. Note that the
critical temperature $T_c$ depends only on $c$ and $\delta \varepsilon$.

Increasing $\lambda$ increases the critical pressure $p_c$. Thus $\Gamma
(T)$ increases its temperature values when $\lambda$ is
increased. Figure~\ref{Fig-3}(c) shows that $\Gamma$ is sensitive to
thermal fluctuations near $T_c$. Figure~\ref{Fig-4} shows that as the
pressure is increased for $v=v_0 + \delta v$, the temperature
decreases. We use our compressible Ising-like model to study the
Ising-nematic phase recently detected in the low-doping regime of
Fe-based superconductors \cite{Rafael}. An electronic nematic phase is
essentially a melted stripe phase \cite{Fradkin155}. Figure~\ref{Fig-4}
shows a description of the nematic phase that fixes the model at $v=v_0
+ \delta v$ (red curve). Here the pressure variation is caused by the
chemical pressure in the system introduced by the doping effect on the
crystal lattice. As the pressure (doping) is varied, the critical point
signature vanishes (see Fig.~\ref{Fig-3}). We obtain the same behavior
shown in Fig.~\ref{Fig-4} (red curve) experimentally for the 122 doped
Fe-based superconductors \cite{budko}. In particular, the thermal
expansion signatures are suppressed upon doping \cite{budko}.

Because there are a variety of ways of fitting the parameters, i.e., a
variety of constants can be varied, we leave fitting the experimental
results reported in Ref.~\cite{budko} to future research. Here we use
our model to simulate the doping effect in single crystals by assuming
there are two different volumes in the melted electronic nematic phase
\cite{Fradkin155}. When the system is doped, the electronic nematic
phase associated with two coexisting volumes (see the figure in
Ref.~\cite{Fradkin155}) is suppressed, and the reported
superconductivity appears, e.g., for Ba(Fe$_{1-x}$Co$_{x}$)$_2$As$_2$
single crystals \cite{budko,ni}.

We have used an energy-volume coupled Ising-like model to calculate the
Gr\"uneisen parameter for the liquid-liquid transition in supercooled
water \cite{claudio}. We find that the behavior of the Gr\"uneisen
parameter diverges near pressure and temperature values that display
anomalously behavior and thus supports the presence of a liquid-liquid
critical point governed by two distinct energy scales. In addition to
exploring the critical behavior of water and its other phases, our model
can also be applied to other systems by adjusting its parameters.

M.\,de S.\,acknowledges financial support from the S\~ao Paulo Research
Foundation -- Fapesp (Grants No.\,2011/22050-4),  National Council of
Technological and Scientific Development -- CNPq (Grants
No.\,302498/2017-6), the Austrian Academy of Science \"OAW for the JESH fellowship and Serdar Sariciftci for the hospitality. The Boston University Center for Polymer Studies is
supported by NSF Grants PHY-1505000, CMMI-1125290, and CHE-1213217, and
by DTRA Grant HDTRA1-14-1-0017.


\begin{thebibliography}{10}

\bibitem{PhysicsToday}
Pablo~G. Debenedett and H.~Eugene Stanley, Physics Today \textbf{56}, 40 (2003).

\bibitem{Gallo}
Paola Gallo, Katrin Amann-Winkel, Charles~Austen Angell, Mikhail~Alexeevich
  Anisimov, Frédéric Caupin, Charusita Chakravarty, Erik Lascaris, Thomas
  Loerting, Athanassios~Zois Panagiotopoulos, John Russo, Jonas~Alexander
  Sellberg, Harry~Eugene Stanley, Hajime Tanaka, Carlos Vega, Limei Xu, and
  Lars Gunnar~Moody Pettersson, Chem. Rev. \texttt{116}, 7463 (2016).

\bibitem{Poole}
Peter~H. Poole, Francesco Sciortino, Ulrich Essmann, and H.~Eugene Stanley,
Nature \textbf{360}, 324 (1992).

\bibitem{Kumar9575}
Pradeep Kumar, S.~V. Buldyrev, S.~R. Becker, P.~H. Poole, F.~W. Starr, and
  H.~E. Stanley, Proc. Natl. Acad. Sci. U.S.A.
  \textbf{104}, 9575 (2007).

\bibitem{Franzese}
Giancarlo Franzese and H~Eugene Stanley, J. Phys.: Condens. Matter \textbf{19}, 205126 (2007).

\bibitem{Gallo1543}
Paola Gallo and H.~Eugene Stanley, Science \textbf{358}, 1543 (2017).

\bibitem{Kim1589}
Kyung~Hwan Kim, Alexander Sp{\"a}h, Harshad Pathak, Fivos Perakis, Daniel
  Mariedahl, Katrin Amann-Winkel, Jonas~A. Sellberg, Jae~Hyuk Lee, Sangsoo Kim,
  Jaehyun Park, Ki~Hyun Nam, Tetsuo Katayama, and Anders Nilsson, Science \textbf{358}, 1589 (2017).

\bibitem{Barto}
Lorenz Bartosch, Mariano de~Souza, and Michael Lang, Phys. Rev. Lett. \textbf{104}, 245701 (2010).

\bibitem{2015}
Mariano de~Souza and Lorenz Bartosch, J. Phys.: Condens. Matter \textbf{27}, 053203 (2015).

\bibitem{EPJ}
Mariano de~Souza, Paulo Menegasso, Ricardo Paupitz, Antonio Seridonio, and
  Roberto~E Lagos, Eur. J. Phys. \textbf{37}, 055105 (2016).

\bibitem{claudio}
Claudio~A. Cerdeiri\~na and H.~Eugene Stanley, Phys. Rev. Lett. \textbf{120}, 120603 (2018).

\bibitem{Cerde}
Claudio~A. Cerdeiri\~na, Gerassimos Orkoulas, and Michael~E. Fisher, Phys. Rev. Lett. \textbf{116}, 040601 (2016).

\bibitem{Cerde1}
Claudio~A. Cerdeiri\~na and Gerassimos Orkoulas, Phys. Rev. E \textbf{95}, 032105 (2017).

\bibitem{huang}
K. Huang, Statistical Mechanics (Wiley, 1987).

\bibitem{garst}
Markus Garst and Achim Rosch, Phys. Rev. B \textbf{72}, 205129 (2005).

\bibitem{Cerio}
F.~Decremps, L.~Belhadi, D.~L. Farber, K.~T. Moore, F.~Occelli, M.~Gauthier,
  A.~Polian, D.~Antonangeli, C.~M. Aracne-Ruddle, and B.~Amadon, Phys. Rev. Lett. \textbf{106}, 065701 (2011).

\bibitem{Review-kappa}
Mariano de~Souza and Lorenz Bartosch, J. Phys.: Condens. Matter \textbf{27}, 053203 (2015).

\bibitem{Zhu}
Lijun Zhu, Markus Garst, Achim Rosch, and Qimiao Si, Phys. Rev. Lett. \textbf{91}, 066404 (2003).

\bibitem{Kuchler}
R.~K\"uchler, N.~Oeschler, P.~Gegenwart, T.~Cichorek, K.~Neumaier, O.~Tegus,
  C.~Geibel, J.~A. Mydosh, F.~Steglich, L.~Zhu, and Q.~Si, Phys. Rev. Lett. \textbf{91}, 066405 (2003).

\bibitem{Roland}
C.~M. Roland, S.~Bair, and R.~Casalini, J. Chem. Phys. \textbf{125}, 124508 (2006).

\bibitem{Cook}
Richard~L. Cook, H.~E. King, and Dennis~G. Peiffer, Phys. Rev. Lett. \textbf{69}, 3072 (1992).

\bibitem{Kadanoff}
Leo~P. Kadanoff, Wolfgang G\"otze, David Hamblen, Robert Hecht, E.~A.~S. Lewis,
  V.~V. Palciauskas, M. Rayl, J.~Swift, D. Aspnes, and J. Kane, Rev. Mod. Phys. \textbf{39}, 395 (1967).

\bibitem{Relaxation-time}
Cecilie R\o{}nne, Per-Olof \AA{}strand, and S\o{}ren~R. Keiding, Phys. Rev. Lett. \textbf{82}, 2888 (1999).

\bibitem{Casalini}
R.~Casalini, U.~Mohanty, and C.~M. Roland, J. Chem. Phys. \textbf{125}, 014505 (2006).

\bibitem{Rafael}
R.~M. Fernandes, A.~V. Chubukov, and J.~Schmalian, Nature Physics \textbf{10}, 97 (2014).

\bibitem{Fradkin155}
Eduardo Fradkin and Steven~A. Kivelson, Science \textbf{327}, 155 (2010).

\bibitem{budko}
S.~L. Bud'ko, N.~Ni, S.~Nandi, G.~M. Schmiedeshoff, and P.~C. Canfield, Phys. Rev. B \textbf{79}, 054525 (2009).

\bibitem{ni}
N.~Ni, M.~E. Tillman, J.~Q. Yan, A.~Kracher, S.~T. Hannahs, S.~L. Bud'ko, and
  P.~C. Canfield, Phys. Rev. B \textbf{78}, 214515 (2008).

\end{thebibliography}
\end{document}